\documentclass{article}

\begin{document}

\begin{center}
\bigskip How to Complement the Description of Physical Universe?

\textit{T.F.Kamalov}

Physics Department\\[0pt]
Moscow State Opened University\\[0pt]
107996 Moscow, 22, P. Korchagin str., Russia\\[0pt]
E-mail: ykamalov@rambler.ru\\[0pt]
\end{center}

\begin{abstract}
Which non-local hidden variables could complement the description
of physical Universe? The model of extended Newtonian dynamics is
presented.
\end{abstract}


\vspace{5mm}

\section{Introduction}

Classical Newtonian mechanics is essentially the simplest way of
mechanical system description with second-order differential
equations, when higher order time derivatives of coordinates can
be neglected. The extended model of Newtonian mechanics with
higher time derivatives of coordinates is based on generalization
of Newton's classical axiomatics onto arbitrary reference systems
(both inertial and non-inertial ones) with body dynamics being
described with higher order differential equations. Newton's Laws,
constituting, from the mathematical viewpoint, the axiomatics of
classical physics, actually postulate the assertion that the
equations describing the dynamics of bodies in non-inertial
systems are second-order differential equations. However, the
actual time-space is almost without exception non-inertial, as it
is almost without exception that there exist (at least weak)
fields, waves, or forces perturbing an ideal inertial system.
Non-inertial nature of the actual time-space is also supported by
observations of the practical astronomy that expansion of the
Universe occurs with an acceleration. In other words, actually any
real reference system is a non-inertial one; and such physical
reality can be described with a differential equation with time
derivatives of coordinates of the order exceeding two, which play
the role of additional variables. This is evidently beyond the
scope of Newtonian axiomatics. Aristotle's physics considered
velocity to be proportional to the applied force, hence the body
dynamics was described by first derivative differential equation.
Newtonian axiomaics postulates inertial reference systems, where a
free body maintains the constant velocity of translational motion.
The body dynamics is described with a second order differential
equation, with acceleration being proportional to force [1]. This
corresponds to the Lagrangian depending on coordinates and their
first derivatives (velocities) of the body, and Euler-Lagrange
equation resulting from the principle of the least action. This
model of the physical reality describes macrocosm fairly good, but
it fails to describe micro particles. Both Newtonian axiomatics
and the Second Law of Newton are invalid in microcosm. Only
averaged values of observable physical quantities yield in the
microcosm the approximate analog of the Second Law of Newton; this
is the so-called Ehrenfest's theorem. The Ehrenfest's equation
yields the averaged, rather than precise, ration between the
second time derivative of coordinate and the force, while to
describe the scatter of quantum observables the probability theory
apparatus is required. As the Newtonian dynamics is restricted to
the second order derivatives, while micro-objects must be
described with equations with additional variables, tending
Planck's constant to zero corresponds to neglecting these
variables. Hence, offering the model of extended Newtonian
dynamics, we consider classical and quantum theories with
additional variables, describing the body dynamics with higher
order differential equations. In our model the Lagrangian shall be
considered depending not only on coordinates and their first time
derivatives, but also on higher-order time derivatives of
coordinates. Classical dynamics of test particle motion with
higher-order time derivatives of coordinates was first described
in 1850 by M.Ostrogradskii [2] and is known as Ostrogradskii's
Canonical Formalism. Being a mathematician, M. Ostrogradskii
considered coordinate systems rather than reference systems. This
is just the case corresponding to a real reference system
comprising both inertial and non-inertial reference systems. In a
general case, the Lagrangian takes on the form

\begin{equation}
L=L(t,q,\dot{q},\ddot{q},...,\dot{q}^{n}).
\end{equation}

\section{Model of Extended Newtonian Dynamics}

Let us consider in more detail this precise description of the dynamics of
body motion, taking into account of real reference systems. To describe the
extended dynamics of a body in an arbitrary coordinate system (corresponding
to any reference system) let us introduce concepts of kinematic state and
kinematic invariant of an arbitrary reference system.

\textbf{Definition}: Kinematic state of a body is set by $n$-th time
derivative of coordinate. The kinematic state of the body is defined
provided the $n$-th time derivative of body coordinate is zero, the $(n-1)$%
-th time derivative of body coordinate being constant. In other words, we
consider the kinematic state of the body defined if $(n-1)$-th time
derivative of body coordinate is finite. Let us note that a reference system
performing harmonic oscillations with respect to an inertial reference
system does not possess any definite kinematic state. Considering the
dynamics of particles in any reference systems, we suggest the following two
postulates.

\textbf{Postulate 1.} Kinematic state of a free body is invariable. This
means that if the $n$-th time derivative of a free body coordinate is zero,
the $(n-1)$-th time derivative of body coordinate is constant. That is,

\begin{equation}
\frac{d^{n}q}{dt^{n}}=0,\frac{d^{n-1}q}{dt^{n-1}}=const.
\end{equation}

In the extended model of dynamics, conversion from a reference system to
another one will be defined as:
\begin{eqnarray}
q^{\prime } &=&q_{0}+\dot{q}t+\frac{1}{2!}\ddot{q}t^{2}+...+\frac{1}{n!}%
\dot {q}^{(n)}t^{n} \\
t^{\prime } &=&t.
\end{eqnarray}

\textbf{Postulate 2.} If the kinematic invariant of a reference system is $n$%
-th time derivative of body coordinate, then the body dynamics is described
with the differential equation of the order $2n$:

\begin{equation}
\alpha _{2n}\dot {q}^{(2n)}+...+\alpha _{0}q=F(t,q,\dot{q},\ddot{q},...,%
\dot {q}^{(n)}).
\end{equation}

This means that the Lagrangian depends on $n$-th time derivative of
coordinate, so variation when applying the least action principle will yield
the order higher by a unity. Therefore, the dynamics of a free body in a
reference system with $n$-th order derivative being invariant shall be
described with a differential equation of the order $2n$. To consider
dynamics of a body with an observer in an arbitrary coordinate system (which
corresponds to the case of any reference system), we apply the least action
principle, varying the action function for $n$-th order kinematic invariant,
we obtain the equation of the order $2n$:

\begin{equation}
\delta S=\delta \int L(t,\dot{q^{\prime }},q^{\prime })dt =\int
\sum_{n=0}^{N}(-1)^{n}\frac{d^{n}}{dt^{n}}\frac{\partial L}{\partial \dot{q}%
^{(n)}}\delta \dot{q}^{(n)}dt=0.
\end{equation}

Then the equation describing the dynamics of a body with $n$-invariant is a $%
2n$-order differential equation, and for the case of irreversible time arrow
we shall retain only even components. Expanding into Taylor's series the
function $q=q(t)$ yields:

\begin{equation}
q=q_{0}+\dot{q}t+\frac{1}{2!}\ddot{q}t^{2}+...+\frac{1}{n!}\dot{q}%
^{(n)}t^{n}.
\end{equation}

It is well known that the kinematic equation in inertial reference systems
of Newtonian physics contains the second time derivative of coordinate, that
is, acceleration:

\begin{equation}
q_{Newton}=q_{0}+vt+\frac{1}{2}at^{2}.
\end{equation}
Let us denote the additional terms with higher derivatives as

\begin{equation}
q_{r}=\frac{1}{3!}\dot{q}^{(3)}t^{3}+...+\frac{1}{n!}\dot{q}^{(n)}t^{n}.
\end{equation}

Then
\begin{equation}
q=q_{newton}+q_{r}.
\end{equation}

In our case, the discrepancy between descriptions of the two models is the
difference between the description of test particles in the model of
extended Newtonian dynamics with Lagrangian $L(t,q,\dot{q},\ddot{q},...,%
\dot {q}^{(n)},...)$ and Newtonian dynamics in inertial reference systems
with the Lagrangian $L(t,q,\dot{q})$:

\begin{equation}
\int [L(t,q,\dot{q},\ddot{q},...,\dot {q}^{(n)})-L(t,q,\dot{q})]dt=h,
\end{equation}
$h$ being the discrepancy (error) between descriptions by the two
models. Comparing this value with the uncertainty of measurement
in inertial reference systems, expressed by the Heisenberg
uncertainty relation, the equation (11) can be rewritten as
\begin{equation}
S(t,q,\dot{q},...\dot{q}^{(n)})-S(t,q,\dot{q})=h.
\end{equation}

In the classical mechanics, in inertial reference systems, the Lagrangian
depends only on the coordinates and their first time derivatives. In the
extended models, in real reference systems, the Lagrangian depends not only
on the coordinates and their first time derivatives, but also on their
higher derivatives. Applying the least action principle [3], we obtain
Euler-Lagrange equation for the extended Newtonian dynamics model:

\begin{equation}
\sum_{n=0}^{N}(-1)^{n}\frac{d^{n}}{dt^{n}}\frac{\partial L}{\partial \dot{q}%
^{(n)}}=0,
\end{equation}

or
\begin{equation}
\frac{\partial L}{\partial q}-\frac{d}{dt}\frac{\partial L}{\partial \dot{q}}%
+\frac{d^{2}}{dt^{2}}\frac{\partial L}{\partial \ddot{q}}-...+(-1)^{N}\frac{%
d^{N}}{dt^{N}}\frac{\partial L}{\partial \dot{q}^{(N)}}=0.
\end{equation}
The Lagrangian will be expressed through quadratic functions of variables:
\begin{equation}
L=kq^{2}-k_{1}\dot{q}^{2}+k_{2}\ddot{q}^{2}-...+(-1)^{\alpha }k_{\alpha }%
\dot{q}^{(\alpha )2}=\sum_{\alpha =0}^{\infty }(-1)^{\alpha }k_{\alpha }\dot{%
q}^{(\alpha )2}.
\end{equation}
For our case, the action function will be:
\begin{equation}
S=q\frac{\partial L}{\partial q}-\dot{q}\frac{\partial L}{\partial \dot{q}}%
+...+(-1)^{\alpha }\dot{q}^{(\alpha )}\frac{\partial \dot{L}^{(\alpha )}}{%
\partial \dot{q}^{(\alpha )}}+...=\sum_{\alpha =0}^{\infty }(-1)^{\alpha }%
\dot{q}^{(\alpha )}\frac{d^{\alpha }}{dt^{\alpha }}\frac{\partial L}{%
\partial \dot{q}^{(\alpha )}}.
\end{equation}

Or
\begin{equation}
S=2kq^{2}-2k_{1}\dot{q}^{2}+2k_{2}\ddot{q}^{2}+...+2k_{\alpha }\dot{q}%
^{(\alpha )2}=2\sum_{\alpha =0}^{\infty }(-1)^{\alpha }k_{\alpha }\dot{q}%
^{(\alpha )2}.
\end{equation}
In the space with curvature,
\begin{equation}
L=\sum_{\alpha =0}^{\infty }(-1)^{\alpha }g_{ik}\dot{q}^{(\alpha )i}\dot{q}%
^{(\alpha )k}.
\end{equation}

Here, instead of Schwarzschild metric
\begin{equation}
ds^{2}=(1-\frac{q_{g}}{q})c^{2}dt^{2}-(1-\frac{q_{g}}{r})dq^{2}-q^{2}d\theta
-q^{2}\sin ^{2}\theta d\phi ^{2},
\end{equation}

we will use the metric
\begin{equation}
ds^{2}=\exp (-\frac{q_{g}}{q})c^{2}dt^{2}-\exp (-\frac{q_{g}}{q}%
)dq^{2}-q^{2}d\theta -q^{2}\sin ^{2}\theta d\phi ^{2},
\end{equation}

where
\begin{equation}
q_{g}=\frac{2GM}{c^{2}},g_{00}=\exp (-\frac{q_{g}}{q}).
\end{equation}%

Introducing the notation

\begin{center}
\begin{equation}
F=\frac{\partial L}{\partial q},p=\frac{\partial L}{\partial
\dot{q}}
\end{equation}%
\begin{equation}
F^{2}=\frac{\partial L}{\partial \ddot{q}},p^{3}=\frac{\partial
L}{\partial \dot{q}^{(3)}}
\end{equation}%
\begin{equation}
F^{4}=\frac{\partial L}{\partial \dot{q}^{(4)}},p^{5}=\frac{\partial L}{%
\partial \dot{q}^{(5)}}
\end{equation}%
.....\\[0pt]
\begin{equation}
F^{2n}=\frac{\partial L}{\partial \dot{q}^{(2n)}},p^{2n+1}=\frac{\partial L}{%
\partial \dot{q}^{(2n+1)}},
\end{equation}
\end{center}

we obtain the description of inertial forces for the extended Newtonian
dynamics model. The value of the resulting force accounting for inertial
forces can be expressed through momentums and their derivatives, expressing
the Second Law of Newton for the extended Newtonian dynamics model:
\begin{equation}
F-\frac{dp}{dt}+\frac{d^{2}}{dt^{2}}(F^{2}-\frac{dp^{3}}{dt})+\frac{d^{4}}{%
dt^{4}}(F^{4}-\frac{dp^{5}}{dt})+...\frac{d^{n}}{dt^{n}}(F^{n}-\frac{dp^{n+1}%
}{dt})=0.
\end{equation}

Expanding the force into Taylor series, we obtain:
\begin{equation}
F(t)=F_{0}+\dot{F}t+\frac{1}{2!}\ddot{F}t^{2}++...
\end{equation}

In other words, (26) can be written as
\begin{equation}
\sum_{n=0}^{\infty }\frac{d^{2n}}{dt^{2n}}(F^{2n}-\frac{d^{2n}p^{2n+1}}{%
dt^{2n}})=0.
\end{equation}
The action function takes on the form
\begin{equation}
S=\sum_{n=0}^{\infty }(-1)^{n}\dot{q}^{(n)}p^{n+1}=\sum_{n=0}^{N}(-1)^{n}%
\dot{q}^{(n)}\frac{\partial L}{\partial \dot{q}^{(n+1)}}.
\end{equation}

The Hamiltonian will be
\begin{equation}
H=\sum_{n=0}^{\infty }\dot{q}^{(n)}p^{n+1}.
\end{equation}

For this case, energy can be expressed as
\begin{equation}
E=\alpha _{0}q^{2}+\alpha _{1}\dot{q}^{2}+\alpha _{2}\ddot{q}^{2}+...+\alpha
_{n}\dot{q}^{(n)2}+...
\end{equation}
Denoting the Appel's energy of acceleration [4] as $Q$, $\alpha _{n}$ being
constant factors, we obtain for kinetic energy and potential energy,
respectively,
\begin{eqnarray}
E &=&V+W+Q  \label{26} \\
V &=&\alpha _{0}q^{2}, \\
W &=&\alpha _{1}\dot{q}^{2} \\
Q &=&\alpha _{2}\ddot{q}^{2}+...+\alpha _{n}\dot{q}^{(n)2}+...
\end{eqnarray}
The Hamilton-Jacobi equation for the action function will take on the form
\begin{equation}
-\frac{\partial S}{\partial t}=\frac{(\nabla S)^{2}}{2m}+V+Q,
\end{equation}
We will call $Q$ the quantum potential. The first addend in (35)
is the so-called Appel's energy of acceleration [4]. The
coordinate derivative of the quantum potential yields the force:
\begin{equation}
F_{Q}=-\frac{\partial Q}{\partial q}.
\end{equation}
The generalized action function will take on the form
\begin{equation}
S=qp+\dot{q}p^{1}+\ddot{q}p^{2}+...
\end{equation}
The velocity is $v=\frac{\partial S}{\partial t}$, and $\frac{\partial v%
}{\partial t}+v\frac{\partial v}{\partial q}=0$, being the continuity
equation of velocity vector. Now, denoting $\frac{\partial S}{\partial \dot{q%
}^{(n)}}=p^{n-1}$, we obtain the equation
\begin{equation}
\frac{dS}{dt}=\frac{\partial S}{\partial t}+\frac{\partial S}{\partial q}%
\dot{q}+\frac{\partial S}{\partial \dot{q}}\ddot{q}+...+\frac{\partial S}{%
\partial \dot{q}^{(n)}}\dot{q}^{(n+1)},
\end{equation}
or
\begin{equation}
\frac{dS}{dt}=\frac{\partial S}{\partial t}+\frac{\partial S}{\partial q}%
\frac{\nabla S}{2m}+p^{1}\ddot{q}+...+p^{n}\dot{q}^{(n+1)}.
\end{equation}
Then
\begin{equation}
\frac{dS}{dt}=\frac{\partial S}{\partial t}+\frac{(\nabla S)^{2}}{2m}+Q.
\end{equation}%
Here, the additional potential $Q$
\begin{equation}
Q=p^{1}\ddot{q}+...+p^{n}\ddot{q}^{(n+1)}.
\end{equation}
Let us complement the equation (2.41) with the continuity equation [5]. In
the first approximation, $Q\approx \alpha _{3}\frac{\nabla ^{2}S}{m^{2}}$
(here, the value of the constant is chosen $\alpha _{3}=\frac{i\hbar m}{2}$.
Hence, in the first approximation we obtain for the function
\begin{equation}
\psi =e^{\frac{i}{\hbar }S},
\end{equation}
the Schroedinger equation\bigskip
\begin{equation}
i\hbar \frac{\partial \psi }{\partial t}=\frac{\hbar ^{2}}{2m}\nabla
^{2}\psi +V\psi.
\end{equation}%
In our case, for free classical particles, we obtain the oscillation
equation:
\begin{equation}
\sum_{n=0}^{N}k_{2n}\dot{q}^{(2n)}=0,
\end{equation}%
or
\begin{equation}
k_{0}q^{2}+k_{2}\ddot{q}^{2}+...+k_{2N}\dot{q}^{(2N)2}=0.
\end{equation}%
The equation (2.46) will take on the form
\begin{equation}
\omega _{0}^{2}q^{2}+\omega _{2}\ddot{q}^{2}+...+\dot{q}^{(2N)2}=0,
\end{equation}%
provided we introduce the notation
\begin{equation}
\omega _{2n}=k_{2n}/k_{2N}.
\end{equation}

\section{Conclusions}

Our case corresponds to Lagrangian $L(t,q,\dot{q},\ddot{q},...,\dot {q}%
^{(n)},...)$, depending on coordinates, velocities and higher time
derivatives, which we call additional variables, extra addends, or hidden
variables. In arbitrary reference systems (including non-inertial ones)
additional variables (addends) appear in the form of higher time derivatives
of coordinates, which complement both classical and quantum physics. We call
these additional addends, or variables, constituting the higher time
derivatives of coordinates, hidden variables or hidden parameters,
complementing the description of particles. It should be noted that these
hidden parameters can be used to complement the quantum description without
violating von Neumann theorem, as this theorem is not applied for non-linear
reference systems, while the extended Newtonian dynamics model assumes
employing any reference systems, including non-linear ones. Comparing the
generalized Hamilton-Jacobi equation
\begin{equation}
\-\frac{\partial S}{\partial t}=\frac{(\nabla S)^{2}}{2m}+V+Q,
\end{equation}%
$Q$ being the additional variables with higher derivatives, with the quantum
Bohm's potential, one can conclude that neglecting higher-order time
derivatives of coordinates brings about incompleteness of physical Universe
description. The coordinate derivative of Q determines the quantum force.
This means that complete description of physical Universe requires
considering differential equations of the order exceeding second;
uncertainty of the position of the particle under investigation shall be
attributed to fluctuations of the reference body and reference system
associated with it. Hence, the differential equation describing this case
shall be of the order exceeding second. In this case, uncertainty of a micro
objects description is follow by incompleteness of the description of the
physical Universe by Newtonian physics, that is, the lack of a complete
description with additional variables in the form of higher time derivatives
of coordinates. The contemporary physics presupposes employment of
predominantly inertial reference systems; however, such a system is very
hard to obtain, as there always exist external perturbative effects, for
example, gravitational forces, fields, or waves. In this case, the
relativity principle enables transfer from the gravitational forces or waves
to inertial forces. For example, if we consider a spaceship with two
observers in different cabins, one can see that this system is non-ideal,
the inertial forces (or pseudo-forces) could constitute additional
parameters here. In this case, superposition of the two distributions
obtained by the observers could yield a non-zero correlation factor, though
each of the two observations has a seemingly random nature. If the fact that
the reference system is non-inertial and hence there exist additional
variables in the form of inertial effects is ignored, then non-local
correlation of seemingly independent observations would seem surprising.
This example could visualize not only the interference of corpuscle
particles, but also the non-local character of quantum correlations when
considering the effects of entanglement.

\end{document}